\documentclass[floatfix,aps,prd,showpacs,eqsecnum]{revtex4}
\usepackage{amsfonts}
\usepackage{amsmath}
\usepackage{amssymb}
\usepackage{natbib}
\usepackage{graphicx}
\usepackage{epstopdf}
\usepackage{subfigure}
\usepackage{psfrag}
\begin{document}
\title{Detecting a Stochastic Gravitational-Wave Background:\\
The Overlap Reduction Function}
\author{Lee Samuel Finn}
\email{LSFinn@PSU.Edu}
\affiliation{Department of Physics, Department of Astronomy and Astrophysics, 
and Center for Gravitational Wave Physics, The Pennsylvania State University, 
State College, Pennsylvania, 
USA 16802-6300}
\author{Shane L. Larson}
\email{s.larson@usu.edu}
\affiliation{Department of Physics,
Utah State University, Logan, Utah, USA 84322-4415}
\author{Joseph D.\ Romano}
\email{joseph.d.romano@gmail.com}
\affiliation{Department of Physics and Astronomy and
Center for Gravitational-Wave Astronomy,
The University of Texas at Brownsville, Brownsville, Texas, USA 78520}
\date{\today}

\begin{abstract}
Detection of a gravitational-wave stochastic background via ground or
space-based gravitational-wave detectors requires the
cross-correlation of the response of two or more independent
detectors. The cross-correlation involves a frequency-dependent factor
--- the so-called \emph{overlap reduction function} or
\emph{Hellings-Downs curve}--- that depends on the relative geometry
of each detector pair: i.e., the detector separations and the relative
orientation of their antenna patterns (beams). An incorrect
formulation of this geometrical factor has appeared 
in the literature, leading to incorrect conclusions regarding the
sensitivity of proposed detectors to a stochastic gravitational-wave
background. To rectify these errors and as a reference for future work
we provide here a complete, first-principles derivation of the overlap
reduction function and assess the nature of the errors associated with
the use of the incorrect expression that has appeared in the
literature. We describe the behavior of the overlap reduction function
in different limiting regimes, and show how the difference between the
correct and incorrect expressions can be understood physically.
\end{abstract}


\maketitle

\section{Introduction}
\label{s:introduction}

The measured response of a single gravitational-wave detector to a
stationary stochastic gravitational-wave signal is indistinguishable
from unidentified instrumental noise.  The gravitational-wave
contribution to the measured response of two or more independent
detectors will, however, be correlated between detector pairs in ways
that other technical noises will not.  The relationship between the
power in a stochastic gravitational-wave background and the
cross-correlated response of a detector pair depends on the response
of the individual detectors and their relative geometry: i.e., their
separation and the relative orientation of their respective detector
antenna patterns, or beams. In the context of ground or space-based
laser interferometric detectors
\cite{luck:2006:sog,barish:1999:lad,waldman:2006:sol,acernese:2006:vs,merkowitz:2007:lis,kawamura:2008:jsg,crowder:2005:ble}
or resonant acoustic gravitational-wave detectors
\cite{cerdonio:2003:agd} this geometrical factor is referred to as the
\emph{overlap reduction function}
\cite{michelson:1987:ods,christensen:1992:msg,flanagan:1993:sol,allen:1997:sgb,allen:1999:dsb};
in the context of pulsar timing arrays \cite{jenet:2006:ubo} or
spacecraft doppler tracking \cite{armstrong:2006:lgw} it is called the
\emph{Hellings-Downs curve} \cite{hellings:1983:ulo}.

Incorrect expressions for the overlap reduction function have appeared
in the recent literature
\cite{cornish:2001:smt,cornish:2001:mgb}
and, with them, incorrect conclusions regarding the sensitivity of
proposed gravitational wave detectors to stochastic gravitational
waves.  These errors have lead to significantly flawed appraisals of
the high-frequency sensitivity of the Big Bang Observer to a
stochastic gravitational wave background, including spurious nulls in
the frequency-dependent detector response and a reduced estimate of
the signal-to-noise ratio as a function of the gravitational-wave
power.  To rectify these errors and as a reference for future work we
provide here a complete, first-principles derivation of the overlap
reduction function and assess the nature and physical interpretation
of the errors associated with the use of the incorrect expression that
has appeared in the literature.

\section{The overlap reduction function}
\label{s:derivation}

The \emph{overlap reduction function} of a pair of gravitational-wave
detectors is the collection of geometric factors, associated with the
relative position and orientation of the detector pair, that appear in
the cross-correlation of the detector pair's response.  Here we derive
an expression for the overlap reduction function by {\em
deconstructing} the cross-correlation, identifying those contributions
that depend only on the radiation and those that depend only on the
detectors, which are then identified as the overlap reduction
function. This approach has the virtue of clearly illustrating the
physical origins of the overlap reduction function and making less
likely mistakes of the kind that may have led to the existing errors
in the literature.

\subsection{Inter-detector cross-correlation}

Consider two gravitational-wave detectors. A stochastic
gravitational-wave ``background'' will manifest itself in a
non-vanishing cross-correlation between the measurements $m(t)$ made
at the two detectors, calculated as a time average over the product of
the measurements:
\begin{align}
C_T(\Delta t,t) &= 
\frac{1}{T}\int_{-T/2}^{T/2} dt'\, m_{1}(t+\Delta t+t')m_2(t+t').
\end{align}
The signature of a stochastic gravitational-wave signal
$h_{ij}(t,\vec{x})$ is just the expectation value of $C_T(\Delta t,t)$
in the presence of $h_{ij}(t,\vec{x})$.  Write the measurement
$m_I(t)$ made at detector $I$ as the sum of a noise contribution
$n_I(t)$ and a signal contribution $r_I(t)$, corresponding to the
detector response to $h_{ij}(t,\vec{x})$
\footnote{Lower case latin indices $i,j,k,l,\cdots$ will denote
spatial components of tensors; upper case latin indices from the
middle of the alphabet $I,J,\cdots$ will label different detectors,
and upper case latin indices from the beginning of the alphabet
$A,A',\cdots$ will denote different gravitational-wave
polarizations.}.  Assume that the noise in each detector is
independent and that there are no non-gravitational-wave effects that
might lead to a correlation in the measurements made at each
detector. Under these assumptions the expectation value of the product
$n_1(t+\Delta t)n_2(t)$ vanishes, implying that the expectation value
of $C_T(\Delta t,t)$ is just the expectation value of the product
$r_{1}(t+\Delta t)r_{2}(t)$:
\begin{equation}
\overline{C}(\Delta t) \equiv 
\overline{C_T(\Delta t,t)} = 
\overline{r_1(t+\Delta t)r_2(t)}\,,
\end{equation}
where overbar denotes expectation value and we have assumed that the
expectation value of $C_T(\Delta t,t)$ is independent of $t$.  This is
equivalent to assuming that the background is {\em stationary}, as we
will describe in more detail in Sec.~\ref{s:statistical_assumptions}.

Turn now to the detector response. Gravitational waves are weak. Even
the most sensitive detectors respond linearly to the local field
$h_{ij}(t,\vec{x})$. Correspondingly we write the detector response as
a convolution, in time \emph{and} space, of an impulse response
function $R_I^{ij}(t,\vec{x})$ with the field $h_{ij}(t,\vec{x})$:
\begin{equation}
r_I(t) 
\equiv
r_I(t,\vec{x}_I)
= 
\int_{-\infty}^\infty d\tau
\int_{R^3} d^3x\>
h_{ij}(t-\tau,\vec{x}_I-\vec{x}) 
R_I^{ij}(\tau,\vec{x})
\,,
\end{equation}
where $\vec{x}_I$ is the spatial location of detector $I$ about which
its response $R_I^{ij}(t,\vec{x})$ is defined.  Causality requires
that $R_I^{ij}(t,\vec{x})$ vanishes outside the future light cone of
$(0,\vec 0)$.  Exploiting the convolution theorem we can also write
\begin{equation}
r_I(t) 
=
(2\pi)^3 
\int_{-\infty}^\infty df
\int_{R^3} d^3k\>
\widetilde{h}_{ij}(f,\vec{k})
\widetilde{R}_I^{ij}(f,\vec{k})
e^{i(2\pi ft-\vec{k}\cdot\vec{x}_I)}
\label{e:rFspace}
\end{equation}
where
\begin{subequations}
\begin{align}
\widetilde{h}_{ij}(f,\vec{k}) 
&=
\frac{1}{(2\pi)^3}
\int_{-\infty}^\infty dt
\int_{R^3} d^3x\>
h_{ij}(t,\vec{x})e^{-i(2\pi ft-\vec{k}\cdot\vec{x})}
\,,
\label{e:htilde}
\\
\widetilde{R}_I^{ij}(f,\vec{k}) 
&= 
\frac{1}{(2\pi)^3}
\int_{-\infty}^\infty dt
\int_{R^3} d^3x\>
R_I^{ij}(t,\vec{x})e^{-i(2\pi ft-\vec{k}\cdot \vec{x})}
\,
\label{e:Rtilde}
\end{align}
\end{subequations}
are the field Fourier modes and detector transfer function. Note that
for real $h_{ij}$ and $R^{ij}$
\begin{subequations}
\begin{align}
\widetilde{h}^*_{ij}(f,\vec{k}) &=\widetilde{h}_{ij}(-f,-\vec{k})\,,
\\
\widetilde{R}_I^{ij}{}^*(f,\vec{k}) &= \widetilde{R}_I^{ij}(-f,-\vec{k})
\,.
\end{align}
\end{subequations}

With the above representations of the detector response, we can
express $\overline{C}(\Delta t)$ in terms of the detector response and
the field:
\begin{subequations}
\begin{align}
\overline{C}(\Delta t) 
&=
\int_{-\infty}^\infty d\tau
\int_{-\infty}^\infty d\tau'
\int_{R^3} d^3x
\int_{R^3} d^3x'\>
\overline{h_{ij}(t+\Delta t-\tau,\vec{x}_1-\vec{x})
h_{kl}(t-\tau',\vec{x}_2-\vec{x}')}
R_1^{ij}(\tau,\vec{x}),R_2^{kl}(\tau',\vec{x}')
\end{align}
or, equivalently,
\begin{align}
\overline{C}(\Delta t) 
&=
(2\pi)^6
\int_{-\infty}^\infty df
\int_{-\infty}^\infty df'\>
\int_{R^3} d^3k
\int_{R^3} d^3k'\>\Big\{e^{2\pi i(f-f')t}
e^{2\pi if \Delta t}
\nonumber\\
&\qquad\qquad
\overline{\tilde h_{ij}(f,\vec{k})\tilde h_{kl}^*(f',\vec{k}')}
\left[\widetilde{R}_1^{ij}(f,\vec{k})
e^{-i \vec{k}\cdot \vec{x}_1}\right]
\left[\widetilde{R}_2^{kl}(f',\vec{k}')
e^{-i\vec{k}'\cdot \vec{x}_2}\right]^*
\Big\}.
\end{align}
\end{subequations}
Note particularly how the detector location and the transfer function
appear together in the combination
$\widetilde{R}^{ij}_I(f,\vec{k})e^{-i\vec{k}\cdot\vec{x}_I}$.  The
form of this combination will be critical when we come to understand
the physical character of the errors made in earlier calculations of
the overlap reduction function.

\subsection{Plane-wave representation of stochastic signal}
\label{sec:planeWave}

Focus attention on gravitational wave fields
$h_{ij}(t,\vec{x})$. These are conveniently represented as a
superposition of plane waves
\begin{align}
h_{ij}(t,\vec{x}) &= 
\int_{-\infty}^\infty df\int_{S^2}d^2\Omega_{\hat{k}}\>
e^{2\pi if(t-\hat{k}\cdot\vec{x})}\mathcal{H}_{A}(f,\hat{k})\mathbf{e}^A_{ij}(\hat{k})
\,,
\label{eq:realH}
\end{align}
where $\hat{k}$ is the unit vector direction of wave propagation, and
$\mathbf{e}^{A}_{ij}(\hat{k})$ are the two orthogonal polarization
tensors,
\begin{subequations}
\begin{align}
2\delta^{AA'}&= \mathbf{e}^A_{ij}(\hat{k}) \mathbf{e}^{A'}_{ij}(\hat{k})\,, 
\\
\mathbf{e}^A_{ij}(\hat{k}) &= \mathbf{e}^{A}_{ji}(\hat{k})
\,.
\end{align}
\end{subequations}
Note that $\mathcal{H}_A^*(f,\hat{k})=\mathcal{H}_A(-f,\hat{k})$ as a
consequence of the reality of $h_{ij}$.  The plane-wave field
amplitudes $\mathcal{H}_A(f,\hat{k})$ are related to the field's
Fourier modes $\tilde h_{ij}(f,\vec k)$ by
\begin{equation}
\mathcal{H}_A(f,\hat{k})\mathbf{e}^A_{ij}(\hat{k}) 
= \left\{\begin{array}{ll}
(2\pi f)^2\,\mathsf{H}_{ij}^{(+)}(f,\hat{k})\,,&f\geq0\\
(2\pi f)^2\,\mathsf{H}_{ij}^{(-)}(f,-\hat{k})\,,&f<0
\end{array}\right.
\end{equation}
where
\begin{equation}
\tilde{h}_{ij}(f,\vec{k}) = 
\mathsf{H}_{ij}^{(+)}(f,\hat{k})\delta(|\vec k|-2\pi f) + 
\mathsf{H}_{ij}^{(-)}(f,\hat{k})\delta(|\vec k|+2\pi f)
\,.
\end{equation}
Here we have introduced separate amplitudes $\mathsf{H}^{(\pm)}_{ij}$
for the positive and negative frequency solutions to the dispersion
relations $|\vec k|^2=(2\pi f)^2$ for a plane wave.

Using expansion (\ref{eq:realH}), we can write the detector response
$r_I(t)$ as
\begin{align}
\label{eq:rjFromMathcalR}
r_I(t) 
=
\int_{-\infty}^\infty df\>
\int_{S^2} d^2\Omega_{\hat{k}}\>
\mathcal{H}_A(f,\hat{k})\mathcal{R}^A_I(f,\hat{k})
e^{2\pi if(t-\hat{k}\cdot\vec{x}_I)}
\,,
\end{align}
where
\begin{align}\label{eq:mathcalR}
\mathcal{R}_I^A(f,\hat{k})
&=
(2\pi)^3
\mathbf{e}^A_{ij}(\hat{k})
\widetilde{R}_I^{ij}(f,2\pi f \hat{k}).
\end{align}
The expectation value $\overline{C}(\Delta t)$ can also be written as
\begin{align}
\overline{C}(\Delta t) 
&= 
\int_{-\infty}^\infty df\int_{-\infty}^\infty df'
\int_{S^2} d^2\Omega_{\hat{k}}
\int_{S^2} d^2\Omega_{\hat{k}'}\>
\Big\{
e^{2\pi i(f-f')t}
e^{2\pi if\Delta t}
\nonumber\\
&\qquad
\overline{\mathcal{H}_A(f,\hat{k})\mathcal{H}^*_{A'}(f',\hat{k}')}
\left[\mathcal{R}_1^A(f,\hat{k})
e^{-2\pi if \hat{k}\cdot \vec{x}_1}\right]
\left[\mathcal{R}_2^{A'}(f',\hat{k}')
e^{-2\pi if'\hat{k}'\cdot \vec{x}_2}\right]^*
\Big\}\,.
\label{e:C(Deltat)}
\end{align}
Again make note of how the detector locations $\vec{x}_1$, $\vec{x}_2$
are associated with the respective transfer functions
$\mathcal{R}_1^A$ and $\mathcal{R}_2^A$.

\subsection{Stationarity, isotropy, and polarization correlations}
\label{s:statistical_assumptions}

The statistical properties of the stochastic signal are encoded in the
expectation values of products of the gravitational field
\begin{equation}
\overline{h_{ij}(t,\vec{x})}
\,,\quad
\overline{h_{ij}(t,\vec{x})h_{kl}(t',\vec{x}')}
\,,\quad
\overline{h_{ij}(t,\vec{x})h_{kl}(t',\vec{x}')h_{mn}(t'',\vec{x}'')}
\,,\cdots
\end{equation}
Without loss of generality, we will assume that any non-zero mean has
been absorbed in the background spacetime, so that
$\overline{h_{ij}(t,\vec{x})}=0$.  Furthermore, for
Gaussian-distributed fields, knowledge of the quadratic correlations
will suffice as all higher-order moments can be constructed from
these.

In our problem we expect that the gravitational wave background is
effectively stationary: i.e., that
$\overline{h_{ij}(t,\vec{x})h_{kl}(t',\vec{x}')}$ depends on $t$ and
$t'$ only through their difference $t-t'$.  In terms of the plane wave
components $\mathcal{H}_A(f,\hat{k})$, this condition becomes
\begin{equation}
\overline{\mathcal{H}_A(f,\hat{k})\mathcal{H}^*_{A'}(f',\hat{k}')} 
= 
\mathsf{H}_{AA'}(f,\hat{k},\hat{k}')
\delta(f-f')
\label{e:stationary}
\,.
\end{equation}
Thus, the different frequency components of a stationary stochastic
background are statistically independent, but they can contribute
differently to the cross-correlated power through the $f$-dependence
in $\mathsf{H}_{AA'}$.  Note that the $\delta(f-f')$ factor in
(\ref{e:stationary}) eliminates the $t$-dependence in
$\overline{C}(\Delta t)$, cf., Eq.~\ref{e:C(Deltat)}.

If the background is \emph{isotropic} --- i.e., the gravitational-wave
specific intensity is independent of the direction of propagation
$\hat{k}$ --- then the most general form of the quadratic expectation
value of the plane wave components $\mathcal{H}_A(f,\hat{k})$ is
\begin{subequations}
\begin{equation}
\overline{\mathcal{H}_A(f,\hat{k})\mathcal{H}^*_{A'}(f',\hat{k}')} 
= 
\mathsf{H}_{AA'}(f,f',\hat{k}\cdot \hat{k}')\,,
\end{equation}
where $\mathsf{H}_{AA'}$ depends on $\hat{k}$ and $\hat{k}'$ only
through the angle between them.  If we further assume that the
components corresponding to different propagation directions are
statistically independent, then
\begin{equation}
\overline{\mathcal{H}_A(f,\hat{k})\mathcal{H}^*_{A'}(f',\hat{k}')} 
= 
\mathsf{H}_{AA'}(f,f')\delta^2(\hat{k},\hat{k}')
\,,
\end{equation}
\end{subequations}
where
$\delta^2(\hat{k},\hat{k}')\equiv\delta(\cos\theta-\cos\theta')\delta(\phi-\phi')$
is the covariant Dirac delta function on the two-sphere.  This latter,
more restrictive, condition is the definition of isotropy for
gravitational-wave stochastic backgrounds typically assumed in the
literature, e.g.,
\cite{christensen:1992:msg,flanagan:1993:sol,allen:1997:sgb,allen:1999:dsb}.

Finally, if the background is \emph{unpolarised}, by which we will
mean that the different polarisation components are statistically
independent and contribute equally to the cross-correlated power, then
\begin{equation}
\overline{\mathcal{H}_A(f,\hat{k})\mathcal{H}^*_{A'}(f',\hat{k}')} 
= 
\mathsf{H}(f,f',\hat{k},\hat{k}')
\delta_{AA'}
\,.
\end{equation}

Putting all these conditions together, we have that an unpolarised,
stationary, isotropic stochastic gravitational-wave background
satisfies
\begin{equation}
\overline{\mathcal{H}_A(f,\hat{k})\mathcal{H}^*_{A'}(f,\hat{k}')}
= 
\mathsf{H}(f)\delta(f-f')\delta^2(\hat{k},\hat{k}')\delta_{AA'}.
\end{equation}
Here $\mathsf{H}(f)$ is a real-valued function proportional to the
gravitational-wave energy density. This is, in turn, directly related
to $\Omega_{\rm gw}(f)$, the ratio of the gravitational-wave energy
density to the cosmological closure density \cite{allen:1999:dsb}:
\begin{equation}
\mathsf{H}(f)=
\frac{3 H_0^2}{32\pi^3}
\frac{\Omega_{\rm gw}(f)}{|f|^3}
\,,
\end{equation}
where $H_0$ is the Hubble expansion rate at the present epoch. 

\subsection{The overlap reduction function}

Combining the results of the previous subsections we find that we can
express the expectation value $\overline{C}(\Delta t)$ of the
inter-detector cross-correlation in the presence of an unpolarised,
stationary, isotropic gravitational-wave background as
\begin{subequations}
\begin{equation}
\overline{C}(\Delta t) = 
\int_{-\infty}^\infty df\>
e^{2\pi if\Delta t} \mathsf{H}(f)\Gamma_{12}(f)
\,,
\end{equation}
where
\begin{align}
\Gamma_{12}(f) &= 
\int_{S^2} d^2 \Omega_{\hat{k}}\>
\mathcal{R}^A_1(f,\hat{k})
\mathcal{R}^{A*}_2(f,\hat{k})
e^{-2\pi if\hat{k}\cdot(\vec{x}_1-\vec{x}_2)}
\,,
\label{eq:or:1}\\
\mathcal{R}^{A}_I(f,\hat{k}) 
&= 
\mathbf{e}^A_{ij}(\hat{k})
\int_{-\infty}^\infty dt
\int_{R^3} d^3x\>
R^{ij}_I(t,\vec{x})
e^{-2\pi if(t-\hat{k}\cdot \vec{x})}
\,.
\label{eq:mathcalR:1}
\end{align}
\end{subequations}
The quantity $\Gamma_{12}(f)$ is the overlap reduction function. It is
often convenient to define a \emph{normalized} overlap reduction
function $\gamma_{12}(f)\propto\Gamma_{12}(f)$ with $\gamma_{12}(0)=1$
for two coincident and coaligned identical detectors. For identical
interferometers with opening angle $\beta$ this leads to the
normalized form
\begin{equation}
\gamma_{12}(f) = \frac{5}{8\pi\sin^2\beta}\Gamma_{12}(f).
\end{equation}

Some symmetry properties of $\gamma_{12}(f)$ follow from immediately
from its definition: in particular,
\begin{subequations}
\begin{align}
\gamma_{12}^*(f) &=\gamma_{21}(f)\,,\\
\gamma_{12}^*(f) &=\gamma_{12}(-f)\,.
\end{align}
\end{subequations}
\section{Discussion}
\label{s:relation}

Summarizing the results of the previous section, the overlap reduction
function normalized for interferometric detectors is
\begin{subequations}\label{eq:or:2}
\begin{align}\label{eq:orGamma}
\gamma_{12}(f) &= \frac{5}{8\pi\sin^2\beta}
\int_{S^2} d^2 \Omega_{\hat k}\>
\mathcal{R}^A_1(f,\hat{k})
\mathcal{R}^{A*}_2(f,\hat{k})
e^{-2\pi if\hat{k}\cdot(\vec{x}_1-\vec{x}_2)}
\,,
\end{align}
where
\begin{align}
\mathcal{R}^{A}_I(f,\hat{k}) 
&= 
\mathbf{e}^A_{ij}(\hat{k})
\int_{-\infty}^\infty dt
\int_{R^3} d^3x\>
R^{ij}_I(t,\vec{x})
e^{-2\pi if(t-\hat{k}\cdot \vec{x})}
\,.
\label{eq:mathcalR:2}
\end{align}
\end{subequations}
Adjusting where necessary for differences in the plane-wave expansion
and detector numbering conventions, 
references \cite{cornish:2001:smt, cornish:2001:mgb}
give the overlap reduction function as
\begin{equation}\label{eq:gammaCL}
\gamma'_{12}(f)
=
\frac{5}{8\pi\sin^2\beta}\int_{S^2} d^2 \Omega_{\hat k}\>
\mathcal{R}_1^A(f,\hat k)\mathcal{R}_2^{A*}(f,\hat k)
e^{2\pi if\hat{k}\cdot(\vec x_1 -\vec x_2)}
\,.
\end{equation}
Comparing this expression with our Eq.~\ref{eq:orGamma} we see that
they differ by the substitution of $e^{2\pi
if\hat{k}\cdot(\vec{x}_1-\vec{x}_2)}$ for $e^{-2\pi
if\hat{k}\cdot(\vec{x}_1-\vec{x}_2)}$.  This difference and when it is
significant can be understood physically; doing so provides the
occasion for a deeper discussion of the overlap reduction function.

\subsection{Detector locations}\label{sec:locations}

At the end of Sec.~\ref{sec:planeWave} we observed that the response
of detector $I$, located at $\vec x_I$, to a field of plane
gravitational waves is given by (cf.~Eq.~\ref{eq:rjFromMathcalR})
\begin{align}
r_I(t) 
=
\int_{-\infty}^\infty df
\int_{S^2} d^2 \Omega_{\hat{k}}\>
\mathcal{H}_A(f,\hat{k})\mathcal{R}^A_I(f,\hat{k})\,
e^{2\pi if(t-\hat{k}\cdot\vec{x}_I)}
\,.
\end{align}
The detector location appears here in the form $e^{-2\pi
if\hat{k}\cdot\vec{x}_I}$.  Referring to Eqs.~\ref{eq:orGamma} and
\ref{eq:gammaCL} for $\gamma_{12}$ and $\gamma'_{12}$ it is clear the
substitution of $\gamma'_{12}$ for $\gamma_{12}$ is equivalent to
simply exchanging the locations of detectors 1 and 2 keeping the rest
of the configuration of the detectors fixed: i.e., $\gamma'_{12}$ is
the overlap reduction function for the detector configuration
consisting of detector 1 at location $\vec{x}_2$ and detector 2 at
location $\vec{x}_1$.  With this understanding we now ask when that
exchange is significant and when it is not.

\subsection{Radiation wavelength and detector separation}\label{sec:smallSep}

An intuitive understanding of $\gamma_{12}(f)$ recognizes that its
behavior in different frequency regimes is governed by several
independent dimensionless parameters that can be created from the
radiation wavelength, the separation between the detectors, and
several intrinsic properties of the detectors as they are represented
in the detector impulse response functions.

Referring to Eq.~\ref{eq:orGamma} we note that when
$f|\vec{x}_1-\vec{x}_2|\ll1$ the exponential term may be replaced by
unity. As this is the only place where the detector separation
appears, in this limit the detector separation plays no role in
determining the value or behavior of $\gamma_{12}(f)$. Defining
\begin{equation}
\delta = f|\vec{x}_1-\vec{x}_2|
\end{equation}
we refer to $\delta\ll1$ as the \emph{small separation limit}. In the
small separation limit, then, the difference between $\gamma_{12}$ and
$\gamma'_{12}$ is negligible under all circumstances.

Now consider the case $\delta\gtrsim1$. As we have observed, the
difference between $\gamma_{12}$ and $\gamma'_{12}$ is the difference
between locating detector 1 at $\vec{x}_1$ or $\vec{x}_2$, and
detector 2 at $\vec{x}_2$ or $\vec{x}_1$.  When the two detectors are
identical in all other aspects, so that $\mathcal{R}^A_1 =
\mathcal{R}^A_2$, this exchange leaves the physical configuration
unchanged and, again, there will be no difference between
$\gamma_{12}$ and $\gamma'_{12}$.

To understand the case $\delta\gtrsim1$ when the two detectors are not
identical we must consider the detector impulse response functions as
they appear in Eqs.~\ref{eq:or:2}.

\subsection{Radiation wavelength and detector impulse response}
\label{sec:smalDet}

The impulse response of a detector has finite support: i.e.,
$R_I^{ij}(t,\vec{x})\simeq0$ for sufficiently large $t>0$ or
$|\vec{x}|$.  A detector doesn't sample the field beyond its physical
extent, so the support in $\vec{x}$ will be on order the detector's
size $\ell$.  Referring to Eq.~\ref{eq:mathcalR:2} it is apparent that
$\mathcal{R}_I^A(f,\hat k)$ depends on $\hat k$ only through
$\mathbf{e}_A^{ij}(\hat k)$ when $f\ell\ll1$:
\begin{equation}
\mathcal{R}_I^A(f,\hat{k})
\simeq
(2\pi)^3\mathbf{e}^A_{ij}(\hat k)
\tilde R_I^{ij}(f,\vec 0)
\,.
\end{equation}
Introducing the parameters
\begin{equation}
\epsilon_I = f\ell_I
\end{equation}
for the two detectors $I=1,2$ we refer to $\epsilon_I\ll1$ as the
\emph{small antenna} limit for detector $I$.

Return now to the difference between $\gamma_{12}$ and $\gamma'_{12}$
when $\delta\gtrsim1$ and the detectors are not identical. When both
$\epsilon_1$ and $\epsilon_2$ are $\ll1$ and noting that
$\mathbf{e}_{ij}^A(-\hat k)=\pm\mathbf{e}_{ij}^A(\hat k)$ we find
\begin{align}
\gamma_{12}(f) 
&\simeq
\frac{5}{8\pi \sin^2\beta}
(2\pi)^6
\tilde R_1^{ij}(f,\vec 0)
\tilde R_2^{kl}{}^*(f,\vec 0)
\int_{S^2} d^2\Omega_{\hat k}\>
\mathbf{e}_{ij}^A(\hat k)
\mathbf{e}_{kl}^A(\hat k)
e^{-2\pi if\hat k\cdot(\vec x_1-\vec x_2)}
\nonumber\\
&=
\frac{5}{8\pi \sin^2\beta}
(2\pi)^6
\tilde R_1^{ij}(f,\vec 0)
\tilde R_2^{kl}{}^*(f,\vec 0)
\int_{S^2} d^2\Omega_{-\hat k}\>
\mathbf{e}_{ij}^A(-\hat k)
\mathbf{e}_{kl}^A(-\hat k)
e^{2\pi if\hat k\cdot(\vec x_1-\vec x_2)}
\nonumber\\
&=
\frac{5}{8\pi \sin^2\beta}
(2\pi)^6
\tilde R_1^{ij}(f,\vec 0)
\tilde R_2^{kl}{}^*(f,\vec 0)
\int_{S^2} d^2\Omega_{\hat k}\>
\mathbf{e}_{ij}^A(\hat k)
\mathbf{e}_{kl}^A(\hat k)
e^{2\pi if\hat k\cdot(\vec x_1-\vec x_2)}
\simeq
\gamma'_{12}(f)
\,;
\end{align}
i.e., in the small antenna limit there is no distinction between
$\gamma_{12}$ and $\gamma'_{12}$.

\subsection{Large separations and large detectors}\label{sec:large}

Finally, consider the case $\delta\gtrsim1$ and, without loss of
generality, $\epsilon_1\gtrsim1$. In this case, the variation of the
field across the spatial extent of detector $1$ is important to the
detector response and, in turn to $\gamma_{12}$. Exchanging the
detector locations changes the relationship between the spatial
\emph{extent and orientation} of detector 1 relative to the location
of detector 2. Correspondingly, in this limit the distinction between
$\gamma_{12}$ and $\gamma'_{12}$ is significant.

\begin{figure}
\begin{center}
\includegraphics[width=5in]{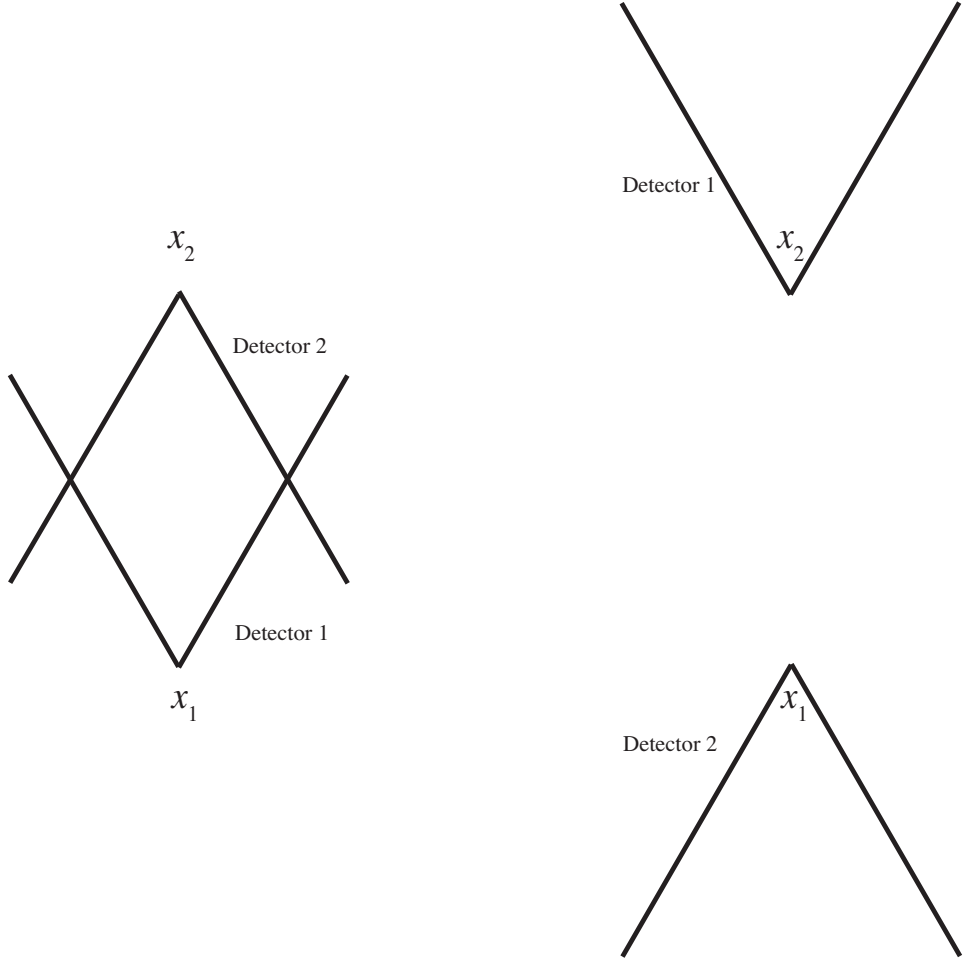}
\caption{The arrangement of the two BBO detectors described
\cite{cornish:2001:smt} and the arrangement actually analyzed when
$\gamma'_{12}$ is substituted for $\gamma_{12}$. Owing to its larger
spatial extent, the system actually studied is much more sensitive to
frequency than the system whose study was intended.}
\label{fig:BBO}
\end{center}
\end{figure}

\begin{figure}
\begin{center}
\includegraphics[width=4in]{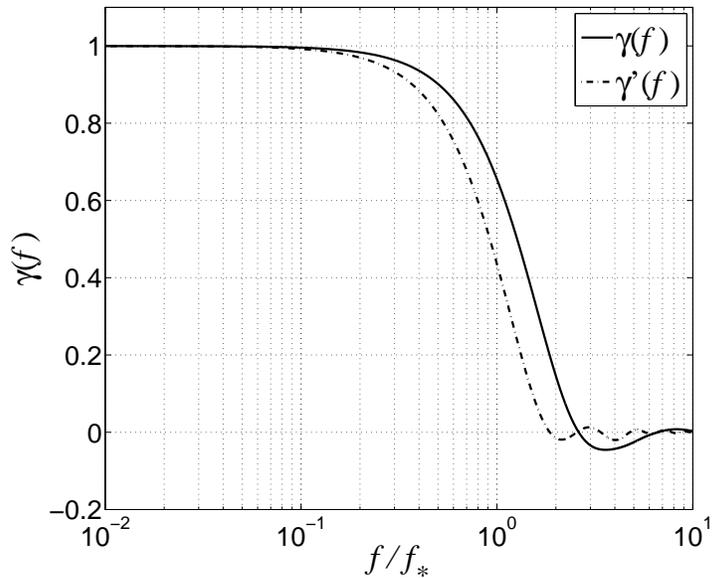}
\caption{Plots of the overlap reduction functions $\gamma_{12}(f)$
(solid line) and $\gamma'_{12}(f)$ (dot-dashed line) for the BBO
configurations in Fig.~\ref{fig:BBO}.}\label{f:overlapBBOComparison}
\end{center}
\end{figure}

\begin{figure}
\begin{center}
\includegraphics[width=4in]{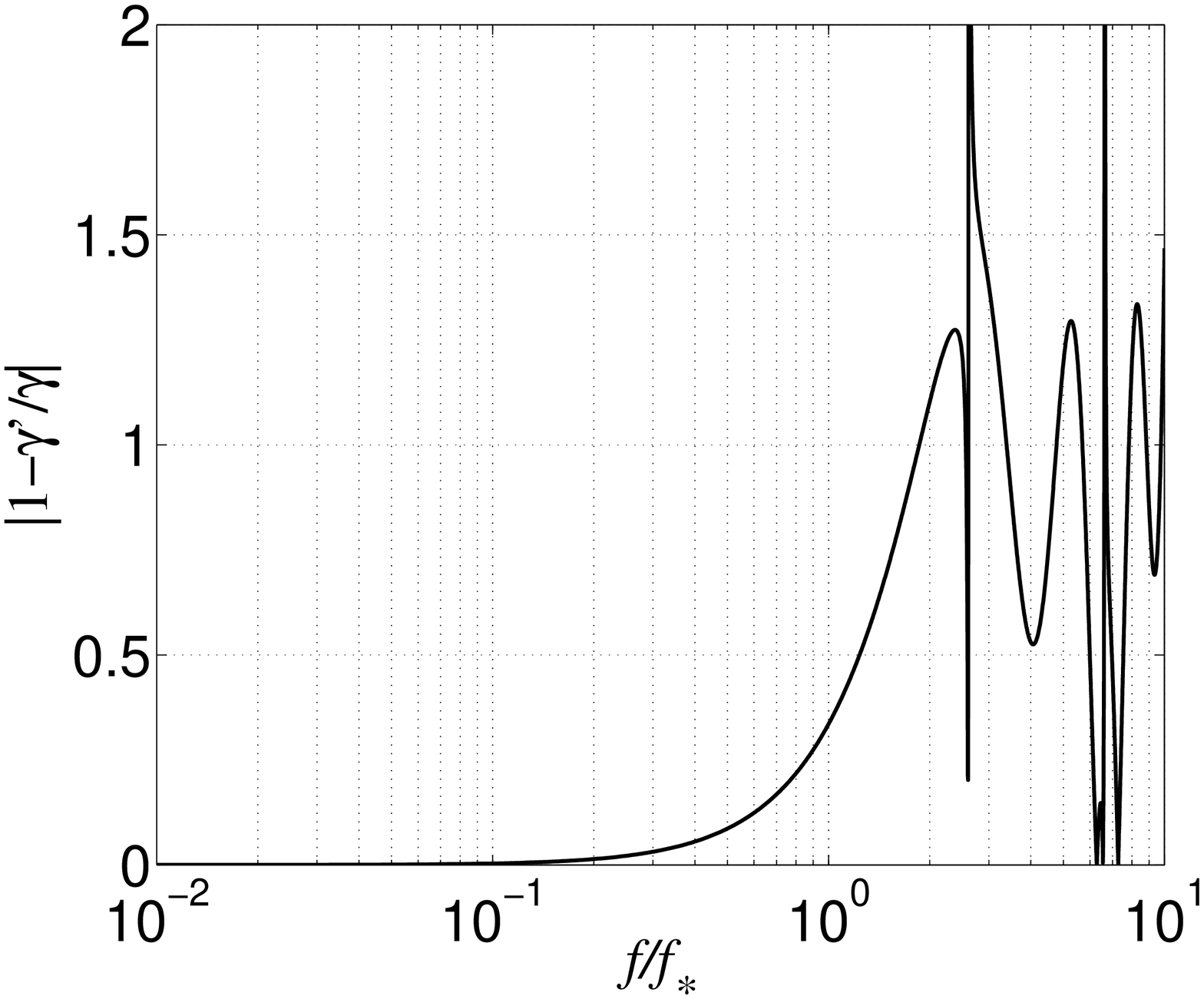}
\caption{Fractional difference between $\gamma_{12}(f)$ and
$\gamma'_{12}(f)$, relative to the correct $\gamma_{12}(f)$ given by
Eq.~(\ref{eq:or:2}).}\label{f:BBOComp2}
\end{center}
\end{figure}

An example of the case $\delta\gtrsim1$,
$\epsilon_1,\,\epsilon_2\gtrsim1$ is the anticipated sensitivity to a
stochastic gravitational-wave background of the Big Bang Observer
(BBO), a space-based follow-on to LISA that has been the subject of
recent study
\cite{cornish:2001:smt,crowder:2005:ble,cutler:2006:bbo,kudoh:2006:dgb,corbin:2006:dcg}. The
principal results of this study reported in the literature
\cite{cornish:2001:smt,cornish:2001:mgb} make use of the incorrect
form of the overlap reduction function, thus mis-estimating this
proposed detector's sensitivity to a stochastic gravitational-wave
signal in the higher frequency regime.
Figure \ref{fig:BBO} illustrates the the physical effect of using
$\gamma'_{12}$ in place of $\gamma_{12}$ when analyzing the
cross-correlation of the two BBO detectors as described in
\cite{cornish:2001:smt}. On the left are the two interferometric
detectors as they are actually arranged in space; on the right are the
effective location and orientation of the detectors when
$\gamma'_{12}$ is used in place of $\gamma_{12}$: i.e., when detector
1 is translated to $\vec{x}_2$ and detector 2 is translated to
$\vec{x}_1$. Under this transformation the spatial extent of the two
detector pairs is much greater than is actually the case;
correspondingly, we expect $\gamma'_{12}$ to be a much more sensitive
function of frequency than $\gamma_{12}$. 

To calculate $\gamma'_{12}$ for comparison with
$\gamma_{12}$,  we need an explicit expression 
for the transfer function of the detectors.
This is derived in \cite[Eqs.~5, 7, 11]{cornish:2001:smt}.
In our notation,
\begin{subequations}
\begin{equation}
{\cal R}^A(f,\hat k)
=
\mathbf{e}^A_{ij}(\hat k)
\frac{1}{2}
\left(u^i u^j {\cal T}(\hat u\cdot\hat k,f) -
      v^i v^j {\cal T}(\hat v\cdot\hat k,f) 
\right)\,,
\end{equation}
where
\begin{equation}
{\cal T}(\hat u\cdot\hat k,f)
=
\frac{1}{2}
\left[
{\rm sinc}\left(\frac{f}{2f_*}(1-\hat u\cdot \hat k)\right)
\exp\left(-i\frac{f}{2f_*}(3+\hat u\cdot\hat k)\right)
+
{\rm sinc}\left(\frac{f}{2f_*}(1+\hat u\cdot \hat k)\right)
\exp\left(-i\frac{f}{2f_*}(1+\hat u\cdot\hat k)\right)
\right]\,.
\end{equation}
\end{subequations}
Here $\hat u$ and $\hat v$ are unit vectors pointing in the 
direction of the detector arms, 
$f_*=c/(2\pi L)$ is the {\em transfer} frequency of the 
detectors (arm length $L=5\times 10^9\ {\rm m}$), and
${\rm sinc}(x)=\sin(x)/x$.
Figure
\ref{f:overlapBBOComparison} compares $\gamma'_{12}$
(cf.~\cite[Fig.~5]{cornish:2001:smt}) and $\gamma_{12}$
(Eq.~\ref{eq:or:2}) for this configuration. As expected $\gamma'_{12}$
decreases much more rapidly with frequency and has its nulls more
closely spaced than those of $\gamma_{12}$.  Figure~\ref{f:BBOComp2}
shows the fractional error
$\left|1-\gamma'_{12}(f)/\gamma_{12}(f)\right|$ as a function of
frequency. The large amplitude spikes in the error occur where nulls
of $\gamma'_{12}$ do not coincide with nulls of $\gamma_{12}$.

It is apparent from Fig.~\ref{f:overlapBBOComparison} that the use of
an incorrect overlap reduction function has, in this case, led to an
underestimate of BBO's sensitivity to a stochastic gravitational wave
background. Estimating the background and detector noise power
spectral density as white, the magnitude of the error is just the
ratio of integrated squared magnitudes $|\gamma_{12}(f)|^2$ and
$|\gamma'_{12}(f)|^2$,
\begin{equation}
\frac{\int df\,|\gamma'_{12}(f)|^2}{\int df\,|\gamma_{12}(f)|^2} = 1-0.28; 
\end{equation}
i.e., the BBO estimates of \cite{cornish:2001:smt,cornish:2001:mgb}
underestimate the sensitivity of BBO by nearly 30\% across the entire
band, and substantially larger if interest is focused on the higher
frequencies. Achieving BBO's goals of detecting the stochastic
gravitational-wave relics of the inflationary epoch depend on the
accurate identification and subtraction of contributions owing to
compact binary systems. Underestimating BBO's response to a
gravitational-wave background leads to an overestimate of the accuracy
required in this identification and subtraction
\cite{cutler:2006:bbo}. Recognizing and correcting the underestimate
in BBO sensitivity thus relaxes the analysis problem associated with
the identification of these foreground sources.

\section{Conclusions}

Detection of a gravitational-wave stochastic background relies on the
cross-correlated response of one or more pairs of gravitational-wave
detectors.  The separation and relative orientation of the two
detectors plays a crucial role in determining the frequency dependent
sensitivity of each detector pair to the stochastic background.
Recent studies
\cite{cornish:2001:smt,cornish:2001:mgb}
of the sensitivity of the Big Bang Observer and related future
generation gravitational-wave detectors have used an incorrect
expression for this geometrical factor.  The errors committed may be
physically interpreted as an exchange in space of the two detectors,
leaving their absolute orientations fixed. In the case of the Big Bang
Observer, this error leads to an approximately 30\% underestimate in
its sensitivity to relic gravitational waves associated with, e.g.,
the inflationary epoch.  Since achieving BBOÕs goals of detecting this
background requires the accurate identification and subtraction of
gravitational-wave foreground contributions from compact binary
systems \cite{cutler:2006:bbo}, this underestimate has lead to
commensurate overestimate of the difficulty of this analysis problem.
Recognizing and correcting this error thus improves, in two ways, the
prospects for the BBO missions main goal as a Big-Bang Observer.

\begin{acknowledgments}
We most gratefully acknowledge the hospitality of the Aspen Center for
Physics and the 2008 Workshop on Gravitational-Wave Astronomy, where
most of the work reported here was done.  JDR also acknowledges
B.~Allen, W.~Anderson, A.~Lazzarini, and J.T.~Whelan for discussions.
Some of the results in this paper were performed using the HEALPix
\cite{gorski:2005:hff} package (\texttt{http://healpix.jpl.nasa.gov}).
The research was supported in part by NSF grant PHY-0555842 awarded to
the University of Texas at Brownsville, NSF grant PHY-0653462 and NASA
grant NNG05GF71G to The Pennsylvania State University, and the Center
for Gravitational Wave Physics, which was supported by the NSF under
cooperative agreement PHY-014375.
\end{acknowledgments}

\end{document}